\documentclass{aa}

\usepackage{txfonts}
\usepackage{natbib}

\usepackage[usenames]{color}

\newcommand{\mps}{m\,s$^{-1}$}

\begin{document}

\title{Large-scale photospheric motions determined from granule tracking and helioseismology from SDO/HMI data}

\author{ Th.~Roudier\inst{1}, M.~\v{S}vanda\inst{2,3}, J.~Ballot,\inst{1}, J.M.~Malherbe\inst{4}, M.~Rieutord,\inst{1}}

\date{Received \today  / Submitted }

\offprints{Th. Roudier}
 
\institute{Institut de Recherche en Astrophysique et Plan\'etologie, Universit\'e de Toulouse, CNRS, UPS, CNES
  14 avenue Edouard Belin, 31400 Toulouse, France
\and Astronomical Institute (v. v. i.), Czech Academy of Sciences, Fri\v{c}ova 298, CZ-25165, Ond\v{r}ejov, Czech
Republic
\and Charles University, Astronomical Institute, Faculty of Mathematics and Physics, V Hole\v{s}ovi\v{c}k\'{a}ch 2, 
CZ-18000, Prague 8,Czech Republic
\and LESIA, Observatoire de Paris, Section de Meudon, 92195 Meudon, France
}

\authorrunning{Roudier et al.}
\titlerunning{Large-scale photospheric motions determined from granule tracking from SDO/HMI data}

\abstract{ Large-scale flows in the Sun play an important role in the dynamo process linked to the solar cycle.  The important large-scale flows  are  the differential rotation and the meridional circulation  with an amplitude of  k\mps{}  and few \mps{}, respectively. These flows also have a cycle-related components, namely the  torsional oscillations.

}{ Our attempt is to determine large-scale plasma flows on the solar surface by deriving horizontal flow velocities using the techniques of solar granule tracking,  dopplergrams, and time--distance helioseismology.

}
{ Coherent structure tracking (CST) and time-distance helioseismology  were used to investigate the  solar differential rotation and meridional  circulation at the solar surface on a 30-day HMI/SDO sequence. The influence of a large sunspot on these large-scale flows with a specific 7-day HMI/SDO  sequence has been also studied.

}
{ The large-scale flows measured by the CST on the solar surface and the same flow determined from the same data with the helioseismology in the first 1~Mm below the surface are in good agreement in amplitude and  direction. The torsional waves are also located at the same latitudes with amplitude of the same order. We are able to measure the meridional circulation correctly using the CST method with only three days of data and after averaging between $\pm 15\degr$ in longitude.
}
{ We conclude that the combination of CST and Doppler velocities allows us to
  detect properly the differential solar rotation and also smaller amplitude flows such as the meridional circulation
  and torsional waves. The results of our methods are in good agreement with helioseismic measurements. 
}


\maketitle

\section{Introduction}

Large-scale motions in the solar convection zone are important elements to understand the
evolution of solar magnetism. Various methods \citep[][]{Pat2010}, based on helioseismology
\citep[][]{giz08,kom14,zhao14}, Doppler velocities measurements \citep[][]{Duv79,hat87},
solar magnetic tracers
\citep[e.g. sunspots, sunspot groups, faculae, bright points; see][]{hat14} or surface feature (e.g. supergranule) tracking  \citep[][]{svanda08,Hat12,Lop17} are
 used  to infer the large-scale solar flows.   Advantages and disadvantages of 
   various techniques are discussed and summarized by \citet[][]{hat14}.

   In particular, magnetic features do not behave as ideal tracers because of their interactions
   with the surrounding plasma \citep[][]{Pat2010}. The use of proxies may induce measurement biases.
   For example, the velocities measured using sunspots or faculae as tracers presumably reflect the properties
   of deeper layers and therefore do not give access to the actual surface plasma motions
   \citep[][]{Java2013,Li13}. It is well establish now that rotation deduced from motion of sunspots is
   systematically faster than that deduced from spectroscopic observations \citep[][]{hat14}. This is partly
because the concentrated magnetic field, which serves as a proxy, is anchored in the
   underlying solar plasma; both the magnetic field and the solar plasma interact strongly in the convective zone, where the plasma $\beta$ 
   parameter, which is the ratio of the plasma pressure to the magnetic pressure $(\beta= 2 \mu P/B^2)$,  is large.
   In this way, the study of the movements of proxies is delicate \citep[][]{sudar17} because
   observational biases can lead to contradictory results, such as in the measurement of the meridional circulation,  the
   amplitude of which is only about 10 m per second.  Helioseismology currently provides the best
   measurement inside the Sun up to near the surface of the plasma flows [from 30 to 1 Mm] \citep[][]{Howe09,Zhao16}

   One of the attempts to determine plasma motions on the surface of the full-disc Sun was made by \citet[][]{svanda08}
 and in a recent work by \citet[][]{Lop17} using local correlation tracking \citep[LCT; ][]{Nov86}. These works
 detect reliable motion of features that are carried by an underlying larger scale velocity field. They also show that a detection of torsional waves and  meridional circulation is possible by applying the LCT method to MDI/SOHO and HMI/SDO Dopplergrams and continuum intensity data, respectively.
 Nevertheless, the LCT method is subject to limitations when applied to whole Sun data. This technique is sensitive
 to the distance to the centre initially because the properties (e.g. size and contrast) of the granules change
 because of projection effects; secondly,  the so-called shrinking effect due to the apparent asymmetry of granulation
 depends on the viewing angle inclined from the vertical, in combination with the expansion of the granules, gives an apparent motion towards the disc centre. \citep[][]{Lop16}.

 Moreover, the LCT is well known for underestimating the amplitude of the velocities by a factor that depends on a particular application and may reach
up to factor two or three \citep[][]{Verma13}.
 
We propose applying the coherent structure tracking \citep[CST;][]{RRRD07,Roud2012} code to follow the proper
motions of solar granules in full-disc HMI/SDO data. By measuring the flow field on the solar surface over a long time period
we obtain a representative description of solar plasma evolution.  The CST allows us to get flow field from covering the
spatial scales from 2.5 Mm up to nearly 85\% of the solar radius.
This method is complementary to helioseismic methods of flow determination, which describes the flows below the solar surface.
  The CST is a granule tracking technique, which allows us to estimate the field direction and amplitude \citep[][]{RRRD07}. The LCT and Fourier local correlation tracking (FLCT) \citep[][]{Welsch2004,Fisher2008} account for both granules and intergranules when cross-correlating continuum images to  estimate the direction and amplitude of the field.
  The principal difference between the LCT
  (FLCT) and CST techniques is that the LCT (FLCT) evaluates the similarity of image subframes at different positions in different
  times. These subframes are required to contain distinct features (granules, supergranule, etc.), but these features are not uniquely
  identified. Thus the subframe displacement is evaluated based on an overall similarity. The LCT  (FLCT) thus provides a smooth
  differentiable estimate of the velocity field. In CST, the code identifies individual features (granules) and tracks these
  individual features coherently through out the image sequence. The resulting velocity field is thus discontinuous and the
  differentiable extension is estimated based on multi-resolution analysis. 

 In this paper, we describe a comparison between velocity fields on the full Sun obtained by various
methods (CST, LCT, FLCT, and time--distance helioseismology). A 30-day sequence of quiet Sun spanning from 16 August to
14 September 2010 is selected. This epoch was particularly selected because the variations of the heliographic
latitude of the solar disc centre ($B_0$) are minimal. This selection allows to us, first, to describe the long-term behaviour of the quiet-Sun 
surface flows and, second, to avoid possible biases introduced by $B_0$ changes, which are mentioned several times in the literature
 \citep[][]{Liang2017,basu2010}. 
We complement this study with the analysis of flows around a large solar spot observed between
12 and 18 April 2016 by HMI/SDO.

We also discuss the data reduction carried out to get an accurate velocity measurements of 
the large-scale flows with a low amplitude. We transform the combination the horizontal components measured via the CST and Doppler observations
to spherical coordinates suitable for a proper description of the solar differential rotation and residual zonal flow, and meridional circulation.
We also study the evolution of the divergence field. Finally, we pay special attention to the effects of a very large sunspot on solar
rotation and meridional circulation. 
 The aim of this work is twofold. First, we use solar granules as passive scalars to follow solar plasma motions
and, second, we obtain quantitative measurements of large-scale movements at the surface of the Sun.

\section{SDO/HMI observations}

\subsection{SDO/HMI observations }

The Helioseismic and Magnetic Imager \citep[HMI: ][]{Scherrer2012,Schou2012} on board the Solar
Dynamics Observatory (SDO) provides uninterrupted observations of the full disc of the Sun. This provides
a unique opportunity to map surface flows on various spatial and temporal scales. We selected
the SDO/HMI continuum intensity data from 16 August to 14 September 2010.  This period was chosen
due to its low solar activity and also to get very small variations of the $B_0$ angle.
The solar differential rotation and meridional circulation are determined from SDO/HMI
continuum intensity and Doppler data taken during this 30-day sequence. 
A second HMI/SDO time sequence from  12 to 18 April 2016 was used to 
describe the effects of the flows around a large sunspot to the solar rotation and meridional circulation.
Both sequences use the original cadence of 45~s and the original pixel size of 0\farcs5. The resolution of the
HMI instrument is 1\farcs0.
 
\subsection{Data analysis }

The flows were measured with two independent methods. 

\subsubsection{Granulation tracking by CST}

 The CST used the solar granules as passive scalars to follow solar plasma motions.
 In order to be suitable for the CST application, the data series of the HMI intensitygrams must
first be prepared. All frames of the sequence were aligned such that the centre of the
solar disc lies exactly on the same pixel in CCD coordinates and the radius of the solar disc was exactly the same 
for all the frames. The reference values for the position of the disc centre and the radius were obtained 
from the first image  (obtained on 16 August 2010 at 00:00:45 UT) of the 30-day series.   Then we performed the granulation  
tracking using the CST code \cite[][]{Roud2012} to reconstruct the projection of the 
photospheric velocity field ($v_{\rm x}$,$v_{\rm y}$) in the plane of the sky (CCD plane) 
\cite[][]{Rincon17}. The application of CST to such a series leads to a sequence of
horizontal velocity field maps in the projection to a sky plane with a temporal resolution
of 30 mins and a spatial resolution of 2.5 Mm (3\farcs5), that is the full-disc velocity map
has a size of 586$\times$586~pixel$^2$. We further removed the $(x, y)$ velocity signal associated with the 
motions of the SDO satellite and Earth's orbital displacements from the CST velocity maps following
the procedure described by \cite{Rincon17}.

The Dopplergrams provide a key piece of information to reconstruct the full vector field. 
The HMI convention is that the line-of-sight (l.o.s.) velocity signal $v_{\rm z}$ is taken as positive when the flow 
is away from the observer, so that the out of plane toward the observer is $v_{\rm dop} =-v_{\rm z}$.
\citep[see Fig 10 in][]{RRPM13}. The processing steps of the Dopplergrams  were to remove Doppler 
shift associated with the proper motion of satellite and Earth's displacement from the 
raw Doppler signal \citep[see][]{Rincon17}. Then in the Doppler data we corrected a 
polynomial radial limb shift function adjusted from ring averages of two hours of data. Since 
the $586^2$~px$^2$  velocity maps obtained from the CST are limited to an effective resolution 
of 2.5~Mm, we then downsampled the $4096^2$~px$^2$ Dopplergrams to the size of CST maps. 
The downsampled  Doppler images were finally averaged over 30 min to match the temporal sampling
of the CST-derived flow maps. 

 The velocity induced by the evolution of $B_0$ in time ($\dfrac{\mathrm{d}B_0}{\mathrm{d}t}$) was corrected
  at the step of the correction of the motions of the satellite.

\subsubsection{Time--distance helioseismology}

The surface flows inferred by CST code were compared to flows in the near-surface layers determined by the time--distance helioseismology
method \citep{Duvall1993}. Time--distance helioseismology is a set of tools that measure and interpret travel
times of seismic waves travelling through the solar interior. Perturbations in the interior cause measurable
shifts in travel times, which can be inverted to learn about the origin of these perturbations.
Plasma flows are very strong perturbers with a clear imprint in the difference travel times, that is in the difference of
travel times of waves propagating between two points in the opposite direction.  

We used a data analysis pipeline described in detail in \cite{Svanda2011}, which was validated against the direct surface
measurements by \cite{SRRBG13}. The time--distance pipeline running at Astronomical Institute of the Czech Academy of Sciences consists of codes for data handling,
filtering (we routinely use both ridge and phase-speed filters), travel-time measurements, and an inversion process. 
The travel times are measured and inverted for flows using plane-parallel approximation in a small
square patch (about 60\degr{} on a side) near the central meridian in a Postel projection. For this study we focussed on 
the travel times of the surface gravity ($f$-) mode with 24-hours averaging in time, which sets 
the random-error levels to 18~\mps{} for horizontal components, and did not investigate the vertical flow component. 

We performed the flow inversions in three patches around the central meridian for a given date centred at latitudes of $-35\degr$, $0\degr$,
and $+35\degr$. The patches were then sewn together, in the course of this procedure the horizontal flows were reprojected to
a Carrington coordinate system. 

The flow maps inferred by our time--distance pipeline have an effective spatial resolution of 10~Mm, which is given by 
the extent of the inversion averaging kernel. 

The horizontal flow maps obtained by the CST code were then compared to time--distance $f$-mode near-surface maps on a pixel-to-pixel 
basis repeating the procedure described in detail in \cite{SRRBG13}. The results of this exercise were almost identical to our earlier comparison, 
that is the correlation coefficient between the CST- and time--distance-derived flows was between $0.7$ and $0.8$ for all 30 days in the series and
both horizontal components. The magnitudes of the flow vectors were also comparable when taking the various effective resolutions into account;
time--distance helioseismology underestimated the flow by some 3--5 per cent. However we noticed a systematic offset of the zonal ($x$) component,
when the time--distance flow estimates were systematically by 30--50~\mps{} larger. No such offset was noticed in the meridional ($y$) component of the flow. We assume that the offset in the zonal component is caused by differences in the data processing pipelines, namely in the tracking procedures. 

\section{Determination of the solar differential rotation}

\subsection{ Solar differential rotation from Dopplergrams}

It is well known that the profile of the differential rotation of the Sun differs by method, data set, and time. However,
there has been rotational profiles published in the literature that are considered a reference. A class of the 
reference profiles was obtained by a spectroscopic method \citep[][] {Pat2010}, where historically the profile 
of,  for example \cite[][]{HH70}, is often used. 

Following the spectroscopic technique, we obtained our reference solar differential rotation profile from the corrected 
sequence of Dopplergrams described above. 

That rotational profile is computed through the relation  
 
\begin{equation}
\Omega(\theta)=\frac{v_{\rm dop}(\theta,\varphi)}{R \cos B_0 \cos\theta \sin\varphi},
\end{equation}
where $\theta$ and $\varphi$  are the latitude and longitude, respectively,
and $R$ the solar radius expressed in km.  
The profile inferred from an average over the 30 days of
observation is shown in Figure~\ref{rotation} (dash-dotted line) and the fit of the polynomial in 
$\sin\theta$ is given, in $\mu$rad\,s$^{-1}$, by

\begin{eqnarray}
\Omega(\theta) &=& 2.87+0.0051 \sin\theta -0.529\sin^2\theta+\nonumber\\
                         &&+ 0.00124 \sin^3\theta -0.395 \sin^4\theta.
\end{eqnarray} 

The fit was performed in longitudes  $\pm 80\degr$ and the corresponding equatorial rotation velocity is $1.998 \pm 0.002 $ k\mps. 

This rotational profiles was used as a reference for comparison of the other techniques. 

 \subsection{ Solar differential rotation from CST and Doppler velocity vector}

One of the first scientific applications of the CST algorithm on SDO/HMI data described in \cite{RRPM13} was  
to determine the solar rotation  from the granule displacements. The horizontal flow $(v_x,v_y)$ measured in the plane-of-the-sky
coordinates by CST code together with  $v_{\rm dop}$ obtained from corresponding Dopplergrams may be transformed 
to spherical velocity components $(v_r, v_\varphi, v_\theta)$. A detailed description of the transform is provided in \cite{RRPM13}, namely in Chapter 5 and Fig.~10. 

In our current study, we further projected $v_\varphi$ and $v_\theta$ to a rectangular $(\varphi,\theta)$ map and, finally, a 
Carrington map is computed from the entire 30-day sequence. Then the mean profile of rotation is obtained by averaging the flow map
in the Carrington coordinates over all longitudes. 
 
The fit of the polynomial in $\sin\theta$ is given, in $\mu$rad\,s$^{-1}$, by

\begin{eqnarray} 
\Omega_\varphi(\theta) &=& 2.62 + 0.0465 \sin\theta -1.70 \sin^2\theta-\nonumber\\
                                      && -0.0177 \sin^3\theta -0.630 \sin^4\theta.
\end{eqnarray}

It seems that the equatorial rotation obtained from $\Omega_\varphi(\theta)$ is 9 per cent smaller than the reference spectroscopic profile computed
from the Dopplergrams. Since spectroscopic profile is our reference, we take the factor of 1.09 to be a correction factor that matches the amplitudes 
of the two techniques that we use further. Figure~\ref{rotation} shows (solid line) the corrected differential rotation profile. We note that this profile is to be compared to a reference spectroscopic profile described in the preceding subsection and plotted by a dash-dotted line in the same figure. In addition to that amplification factor, the two rotational curves show a very good correspondence in the northern region, but a slight 
underestimation (around 4 m/s) of the solar rotation in the southern region.

 The observed asymmetry about the equator seen with the CST is in great part due to the high $B_0$ angle around  $+7\degr$,
  which implies a larger projection effect and lower amplitude velocity in the south. Such asymmetry was not detected up to
  $80\degr$ in our first SDO data analysis (see fig 6 of \cite{RRPM13}), on 10 December  2011, where $B_0=-0.28\degr $
  was very small, reducing the projection effects. The correction for $B_0$ evolution described in section 2.2.1 does not take
into account the foreshortening of the granules, which introduces a systematic
bias in the velocity field. Both components of horizontal velocity are affected by this issue and are the main cause of the observed asymmetry.
However, in the equatorial patch this effect is small and an apparently observed asymmetry about the equator is consistent within various
methods (e.g. not only CST, but also time--distance).

\subsection{Solar differential rotation from helioseismology}

Similar to the CST method, the differential rotation profile was derived also for the time--distance $f$-mode flow. Since during the stitching of the tiles
investigated separately at latitudes $0$ and $\pm 35$\degr{} the flow was already transformed to the Carrington coordinates, we only needed to averaged a synoptic map over longitudes and fit the polynomial in $\sin\theta$. 

The fit, again in $\mu$rad\,s$^{-1}$, is
\begin{eqnarray}
\Omega(\theta) &=& 2.84 +0.00772\sin\theta -0.192\sin^2\theta-\nonumber\\
&&-0.0415\sin^3\theta -0.805\sin^4\theta.
\end{eqnarray}

Such a profile gives an equatorial velocity of $1.977$~k\mps, which is in good agreement with a reference equatorial rotation speed inferred from the spectroscopic method. 

The helioseismic profile averaged over the 30~d of observation is again plotted in Figure~\ref{rotation}. Compared to other techniques, the agreement is very good in equatorial regions and deviates significantly towards larger latitudes; the helioseismic profile is more solid than the other techniques. At 50\degr{} latitude the difference in speeds is almost 200~\mps. This is probably due to the non-linear effects, pointed out already by \cite{Jackiewicz2007}. Flows with magnitudes larger than some 150~\mps{} are underestimated by linear helioseismic approach, as beyond this limit even the velocity becomes a strong perturber and the linearity assumption is violated. The zonal flows at large latitudes reach the amplitudes of several hundred \mps{} compared to the tracking rate, so naturally the non-linearity steps in.

\subsection{Solar differential rotation from divergence maps}

The horizontal velocities $v_x$ and $v_y$  from CST code allow us to access directly the divergence field over the
full Sun with a time step of 30 min (Figure~\ref{div}) in the  30-day sequence. This field reveals
the superganular pattern, which contributes to a diffusion of the magnetic field over the solar surface. To study the 
field, we first remap the divergence maps using a Sanson-Flamsteed (sinusoidal) projection following the transform
\begin{equation}
X=\varphi \cos\theta\quad {\rm and}\quad  Y=\theta.  
\end{equation}
The Sanson-Flamsteed projection conserves the areas, which is a necessary requirement for the following steps \citep{SKS06}. 

From the sequence of divergence maps we infer the apparent velocities by two methods: LCT \citep[][]{Nov86} and
FLCT \citep[][]{Welsch2004,Fisher2008}. Both methods in essence track the features (divergence centres that mark the location of supergranular cells) in a sequence of frames and measure the continuous displacement field that is in the end converted to horizontal velocities
in the $(X,Y)$ coordinates. 

The inferred velocities $v_X$ and $v_Y$ are then converted to cylindrical coordinates $(\varphi,\theta$) on a sphere using a transformation
\begin{eqnarray}
v_\varphi &=& R \left( v_X +  \varphi v_Y \sin\theta \right),\\
v_\theta &=& R v_Y
\end{eqnarray}
 with
\begin{eqnarray}
v_X&=&dX/dt,\\
v_Y&=&dY/dt
\end{eqnarray}   

\begin{figure}
\centering
\includegraphics[width=9cm]{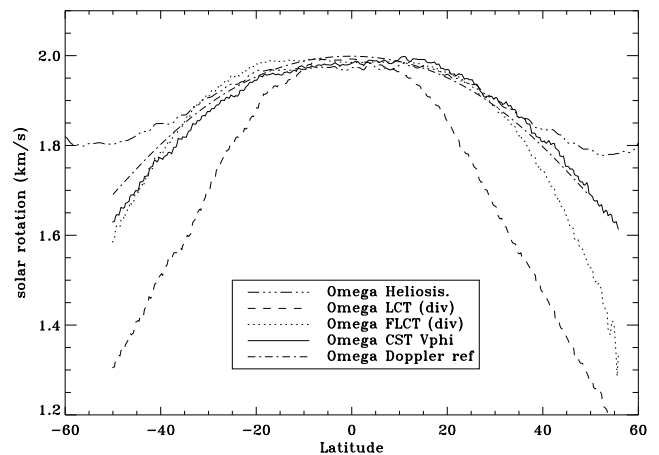}
 \caption[]{Solar zonal rotation $\Omega(\theta)$ expressed in k\mps.}
\label{rotation}
\end{figure}

As it was carried out for the other methods, $(v_\varphi,v_\theta)$ velocities from both LCT and FLCT are projected to Carrington maps
for all the 30 days of the sequence. From the synoptic maps we again obtained the averaged flow profiles as a function of the 
latitude by averaging over longitudes. The differential rotation
 from LCT and FLCT methods using divergence field as tracers are shown in Figure~\ref{rotation}   (dashed line for LCT and
 dotted for FLCT). Both tracking methods are known to underestimate the velocities, thus we multiplied these methods by an empirically
obtained factors of 1.15 for LCT and 1.065 for FLCT, respectively.
 
Obviously, the differential rotation curve from LCT differs significantly from the other methods outside the equatorial 
zone, where a rotation curve is more differential. The FLCT curve agrees very well with a Doppler reference up to
$\pm 35\degr$, farther from the equator the velocities are underestimated, particularly in the northern hemisphere.
The LCT and FLCT do not measure the same flows as the Doppler measurement does. The LCT and FLCT are sensitive to the
  divergence and inter-divergence horizontal displacement evolution, whereas the Doppler measurement is affected by its radial location relative to the disc centre. Indeed the Doppler velocity at the disc centre represents vertical flow at 100 km of altitude and away from
  that location Doppler velocities are sensitive to the horizontal flows at around 200 km of altitude \citep{FCS2011}. This
  effect is partly corrected by the limb-shift correction. All the methods (LCT, FLCT, Doppler, and CST) are affected by projection effects
  that mix the granular flows outside the disc centre.

\begin{figure}
\centering
\includegraphics[width=9cm]{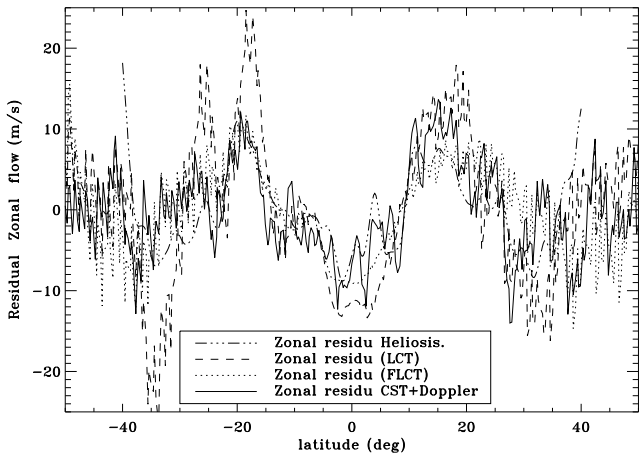}
 \caption[]{Solar residual rotation expressed in \mps.}
\label{residu}
\end{figure}

\begin{figure}
\centering
\includegraphics[width=9cm]{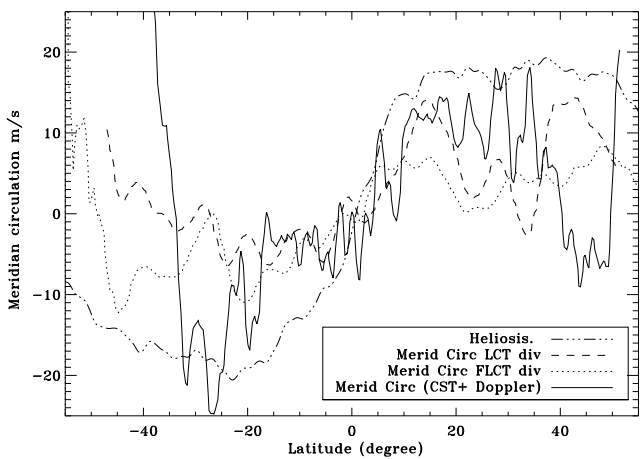}
 \caption[]{Meridian circulation expressed in \mps.}
\label{circ}
\end{figure}

\begin{figure}
\centering
\includegraphics[width=9cm]{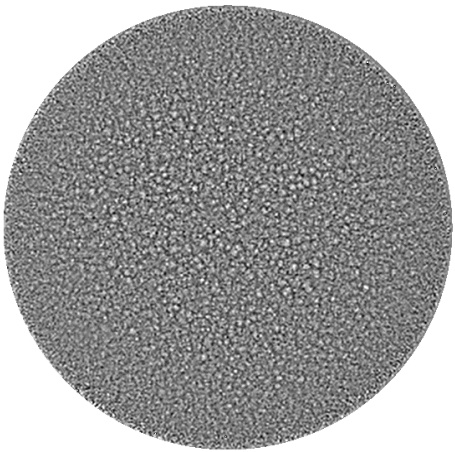}
 \caption[]{Divergence map computed from the CST velocities $v_x$ and $v_y$.}
\label{div}
\end{figure}

\section{Residual zonal flow}
 
The solar rotation has a time-varying component, the torsional oscillations \cite[][]{How1980,Kosov16,Zhao16},
which can be measured via the residual zonal flow. This component has
a very low amplitude (a few \mps) compared to the magnitude of the solar rotation (almost 2 k\mps{} at the equator)
and needs to be revealed by  applying a sufficiently large time averaging window.  

In order to obtain the residual flow for a 30-day average, we first fit the averaged zonal flow profile (the differential rotation)
by a fourth-degree polynomial in $\sin\theta$ and then subtract this fit from the averaged flow profile. Figure~\ref{residu} shows the
residual zonal flow inferred by various methods of measurement. All the plots show the same behaviour,
in which maxima are found around $15\degr$ to $20\degr$ in the northern region (positive latitudes) and around
$-20\degr$ in the south region (negative latitudes). The central depression is found at the equator.
The amplitude at the maxima are between $\pm10$~\mps. These values, the latitude of the maxima, and their
amplitude are in good agreement, at the same date as that found in the literature
\cite[e.g.][]{zhao14,kom14,giz08}. The residual zonal flow obtained by the LCT exhibits a larger amplitude
of the fluctuation in the southern region relative to the other curves. 

 The exact appearance of the torsional oscillation pattern is quite sensitive to the background term that is subtracted
  (see Figures 24 and 25 in \citep[][]{Howe09}). Two methods are used to reach the detection the  torsional oscillation. The first method consists of the subtraction of the rotation profile measured in the quiet (minimum) periods from the profiles in the active
  periods. Both profiles are averaged over longer times (typically many Carrington rotations)\citep[][]{Howe2005,Howe2006}. The second
  method, which is largely consistent with that above, is the subtraction of a smooth fit (usually a polynomial in the sine
  latitude) from the general profile. The later method is common for time--distance helioseismic results and enables us to study time evolution of the profile (see e.g. \citep[][]{ Zhao2004}).
We conclude that the combination of CST and Doppler velocities allows us to
detect torsional waves  properly. We also note that the results of our methods is in good
agreement with helioseismic measurements.

\section{Meridional circulation flow}
 
 The meridional circulation is a long-lived photospheric flow amplitude that is around 10 to 20 \mps{} and 
 requires the averaging of measured flows in time. The profile of the meridional flow as a function of latitude is obtained by averaging the meridional
$v_\theta$ flow component over longitudes, which uses  an analogous method as for the determination of the 
differential rotation, only applied to the other component of the horizontal flow.

 Figure~\ref{circ}  shows the profiles of surface meridional circulation obtained from the various methods discussed
  above. For all the curves we note a detection of the meridional circulation between $-25$\degr{} and $+15$\degr{} but with a
  smaller amplitude for the LCT and FLCT methods. The trend seen in the combination of CST horizontal velocities and the Doppler
  component agrees very well with the trend of the helioseismic determination, the amplitudes of both are more or less similar
  around $\pm15$~\mps, however the slope seems about twice as steep in case of the helioseismic inference.
  The velocities determined with CST and Doppler data however do not allow reliable measurements beyond
 $-40\degr$. This is because of the $B_0$ angle (between $6.70\degr$ to $7.22 \degr$), which induces a large
projection effect (foreshortening) of the solar granulation at high latitudes in the southern region.
Owing to the foreshortening effect, which is symmetrical about the disc centre, averaging over longitude from the
  central meridian is limited around $\pm45$\degr.
  It is difficult to directly compare anything other than the trends in the case of data set we have at our disposal.
  For instance, the intrinsic variability of the two profiles has different typical spatial scales, where the profile obtained
  from the combination of CST and Doppler velocities varies much faster with latitude than the helioseismic profile.
  A proper study would require a much larger sample of velocities and is beyond the scope of a current study. 

By integrating 30 days  of data, we are able to recover the surface meridional flow around the equator with the various techniques
 we mutually compare. The meridional flows are consistent between the methods and clearly show motions towards the poles in both hemispheres.

\begin{figure}
\centering
\includegraphics[width=9cm]{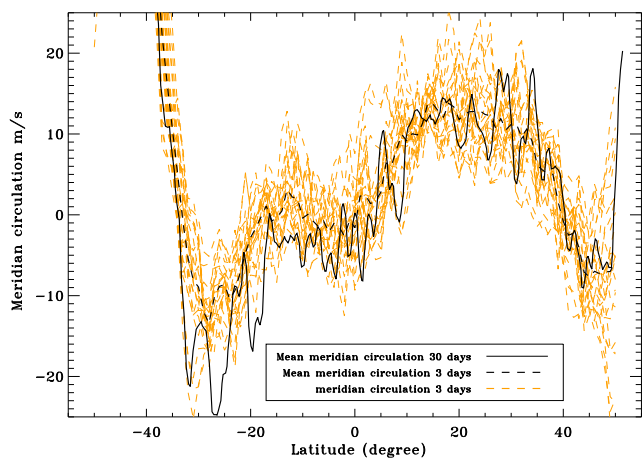}
\caption[]{Meridional circulation  measured with a moving temporal window of 3 days during the 30-day sequence
  (orange dashed line), meridional circulation issued from the temporal averaging of 30 days (solid line) and
  the average of all the 3-day averages  (dashed lines). }
\label{days3}
\end{figure}

  Owing to its very small amplitude, the reliable measurement of the  meridional circulation flow requires an averaging of a  long time sequence 
  \cite[][]{Hath96,giz08,Ulrich2010,kom14,Zhao13,Kosov16,Zhao16}. We further focus on measurements obtained by a combination of 
CST horizontal velocities and Doppler velocities. Our goal is to determine how short the temporal window can be
to obtain a reasonable signal-to-noise ratio to be able study its evolution during the solar cycle. In
  the parameter space of a temporal window averaging and longitude averaging domain centred 
  on the central meridian, we tested various configurations of that parameter. The best compromise to
  minimise the temporal averaging is found for a temporal window of 3 days and a longitudinal averaging between
  $\pm 15 \degr$. 

We evaluate the optimal parameter selection by comparing the short-term and longitude span with a 30-day all-longitudes average. For the
parameter selection described above the short-term and longitude span curves differ from the long-term all-longitudes average with a standard
deviation of 5~\mps{}, which is a reasonable value given the fact that the amplitude of the meridional flow is about 15~\mps. In Figure~\ref{days3} we plot the long-term all-longitudes average (solid line) together with all 3-day averages along a patch in longitudes of $\pm 15 \degr$ around the central meridian (orange dashed lines), and the average of all the 3-day averages  (dashed lines). We note that when the plotted 3-day averages are averaged
together, they do not agree exactly with the 30-day reference, however the differences are small (around 2--4\mps{} in the south part). That is because the 3-day averages are computed directly from $v_\varphi$ maps, whereas the 30-day average is computed from the Carrington synoptic maps.

\begin{figure}
\centering
\includegraphics[width=9cm]{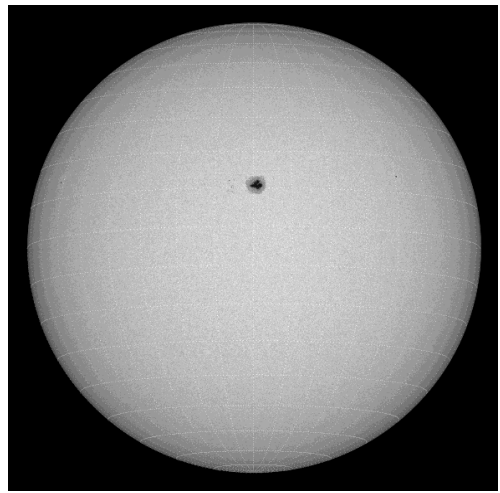}
\caption[]{Context image of the sunspot at the central meridian on 13 April 2016 at 23:00~UT.}
\label{sunspot}
\end{figure}

\begin{figure}
\centering
\includegraphics[width=9cm]{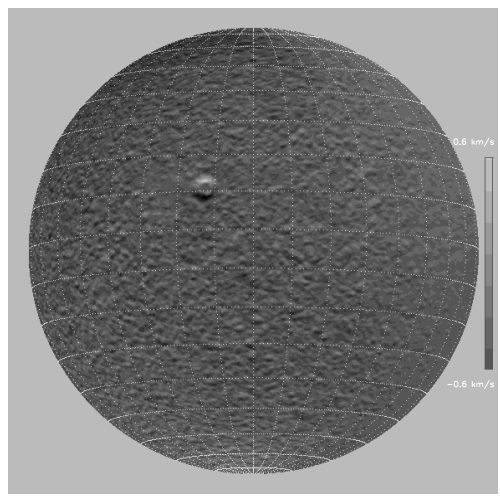}
\caption[]{Image of $v_\theta$ on 12 April 2016. Only quiet-Sun regions occupy the central meridian.}
\label{quiet}
\end{figure}

\begin{figure}
\centering
\includegraphics[width=9cm]{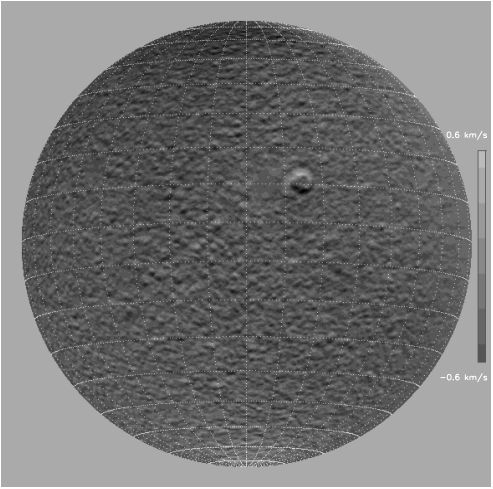}
\caption[]{Image of $v_\theta$ on 14 April 2016. The plage region is located at the central meridian.}
\label{plage}
\end{figure}

\begin{figure}
\centering
\includegraphics[width=9cm]{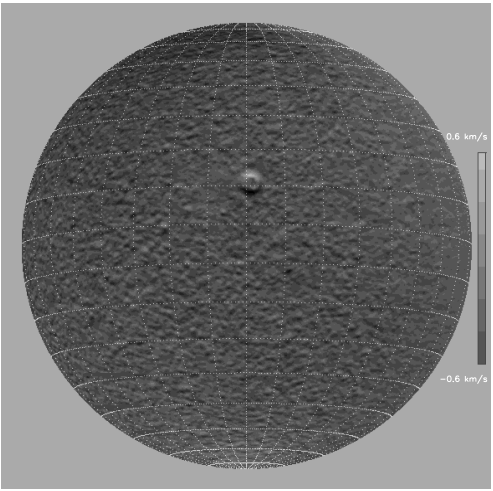}
\caption[]{Image of $v_\theta$ on 13 April 2016. The sunspot is positioned at the central meridian.}
\label{active}
\end{figure}

\begin{figure}
\centering
\includegraphics[width=9cm]{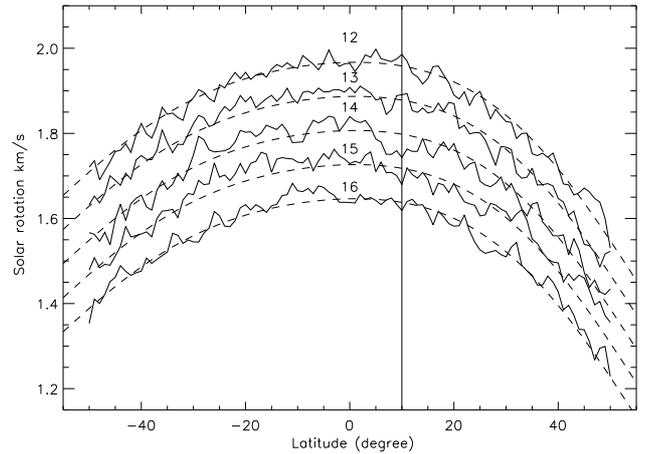}
\caption[]{Solar zonal rotation $\Omega(\theta)$ expressed in k\mps{}. For clarity the plots are
  shifted down in amplitude from date to date.}
\label{zonal}
\end{figure}

\begin{figure}
\centering
\includegraphics[width=9cm]{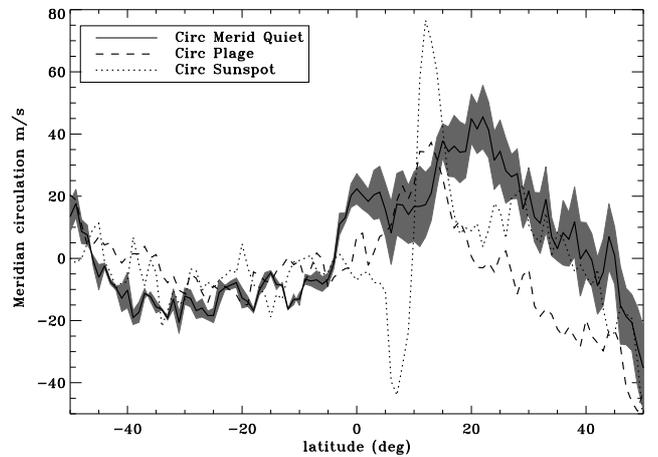}
\caption[]{Meridian circulation expressed in \mps{} in a quiet Sun  and its extrema (grey shaded), sunspot, and plage regions.}
\label{circ_all}
\end{figure}

\section{Effects of a stand-alone sunspot on large-scale flows}

The presence of small sunspots and facula plages on the Sun during the 
30-day series did not allow us to study the effect of such magnetic structures 
on the large-scale flows. In order to test such an influence, we selected a complementary series (12 to 18 April 2016) 
with a very large sunspot (NOAA 12529, Figure~\ref{sunspot}). This isolated sunspot crossed the central meridian
at latitude of $10\degr$ on 13 April 2016 and had the projection-corrected area of 960 millionths of solar hemisphere according to the
archive of Heliophysical Observatory, Debrecen.

We attempted to study the effects of the presence of magnetised plasma and thus we studied separately three different
situations: a quiet-Sun region, the flow field around a sunspot, and a flow field around the plage region following the
spot.  In this complementary study, we focussed on the velocities determined by the combination of horizontal CST and Doppler velocities only.
To capture the characteristic behaviour and avoid possible projection effects, we always studied only a patch
around the central meridian and used the solar rotation that carried the regions of interest into this particular patch during the entire
sequence.

In this way, all the regions, including quiet, plage, and sunspot regions, were derotated to the central meridian.
Then, we applied an averaging in longitude between $-2\degr$ and  $12\degr$  for the quiet Sun (Figure~\ref{quiet}) and
plage (Figure~\ref{plage}), and  between $-7\degr$ and  $7\degr$ for the sunspot (Figure~\ref{active}), to have the averaged
patch well centred on the specified structure.

Figure~\ref{zonal} shows the daily averages of the zonal velocity for a sequence of 5 days.
The dashed line represents the reference of the quiet-Sun rotation obtained for 12 April 2016, for which we fit the
fourth degree polynomial in $\sin\theta$ as usual. This fit is given in $\mu$rad\,s$^{-1}$ by the formula
\begin{eqnarray}
\Omega(\theta) &=& 2.75 +0.0173 \sin\theta -0.431 \sin^2\theta+\nonumber\\
&&+ 0.280 \sin^3\theta -0.640 \sin^4\theta,
\end{eqnarray}
giving an equatorial velocity of $1.91 \pm 0.005 $ k\mps. In Figure~\ref{zonal} the amplitude is normalised to the
direct Doppler  solar rotation determination of the same day, following the procedure introduced in the previous sections. 
To avoid overlapping of the curves in Figure ~\ref{zonal}, we introduced an artificial shift in amplitude date to date. 

At the sunspot latitude ($10\degr$) we observe a depression on 14 April, which is related to the 
presence plage and not to the sunspot, which was at the central meridian a day earlier.  
 That depression is related to a lowered solar rotation which in fact is due to the proper motion 
of the following  faculae plage moving eastward with a mean velocity of  80~\mps. This separation velocity 
amplitude is compatible  with the length of the evolutionary phases \citep[larger than 5 days; see][]{Verma2016}.
 The solar differential rotation does not appear to be sensitive to large-scale sunspot flows because they are largely radially symmetric, thus
having opposite directions eastward and westward from the spot and thus practically cancelling each other out. 

Figure~\ref{circ_all} shows the mean meridional circulation for the quiet-Sun, sunspot, and plage regions which are
centred onto the central meridian similar to the differential rotation describe above.
 In order to get the dispersion of the velocities  in  the quiet Sun, which is our reference region, the temporal window was
  fixed to 3 days (like in Fig. 5) and 5 days for plage and sunspot region. The quiet Sun exhibits the classic decreasing
behaviour between the northern and southern regions and amplitude in the quiet Sun lies between $-25$ and $30$~\mps.
 The grey shaded area in Figure~\ref{circ_all} represents the extrema of the meridional circulation for the quiet Sun. The
  larger dispersion of the meridional circulation in the northern region shows the sensitivity of that signal to the projection effects
  ($B_0$=$-5.72$\degr). The amplitude is larger than determined for the 30-day series (Fig. 5) probably due to the different cycle
of activity and projection effects.

The meridional circulation appears to be affected by the sunspot region in which the moat flow modifies greatly the
large-scale flow with amplitude up to 100~\mps. So during the sunspot life (few days or more) the diffusion
of the magnetic elements  in the meridional direction are probably affected and must be taken into account
in the diffusion model of magnetic field on the Sun surface \cite[][]{Cosset17}. 

In case of the plage region we note a similar behaviour to the quiet region in the southern region, but a smaller amplitude
in the northern region. We do not know  the origin of that amplitude reduction. We question if it is due to the lower amplitude in the plage region,
which does not allow the transfer of the regular flow to higher latitudes or to a pure observational coincidence.
New observations are required to conclude that point. With only a few day average it is not possible to reliably 
detect the torsional oscillations in the residual flow as the amplitude of the variations of such residual flow is
about 10~\mps, which is larger that the expected magnitude of torsional oscillation of 4~\mps.

\begin{figure}
\centering
\includegraphics[width=9cm]{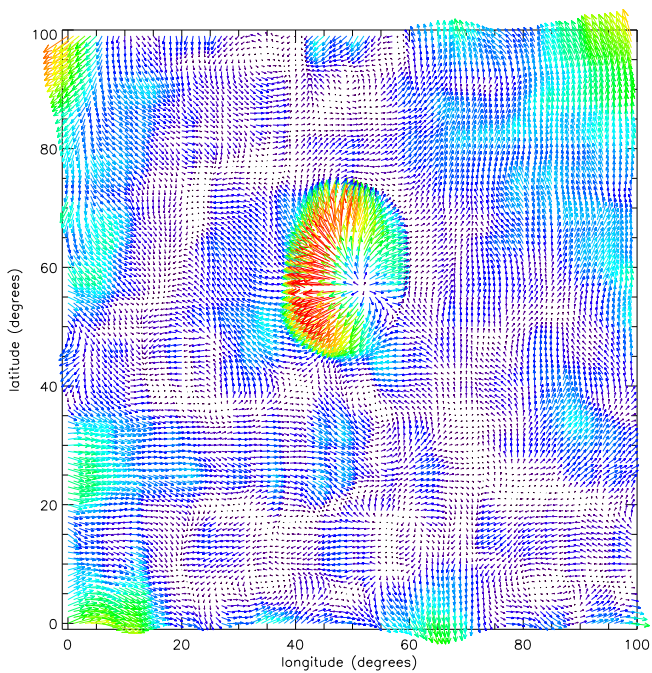}
\caption[]{Flow fields around the sunspot averaged over 5 days and 7 degrees of spatial window. }
\label{flow_sunspot}
\end{figure}

Following \cite{Lop17} we measured the large-scale flow around the large sunspot.  We used the horizontal flows
measured in our 5-day sequence, tracked the sunspot location over the whole series, and averaged all flow maps
such that they are co-spatial in the location of the spot. Figure~\ref{flow_sunspot} shows the flows in which the sunspot moat diverging motions are  
clearly visible. With that temporal (5 day) and spatial $6\degr$ LCT window,  we did not observe the large-scale 
converging flow to the sunspot observed by \cite[][]{Lop17}. The amplitude of the measured flows are around 30 m/s
except in the moat where the amplitude goes up to 149 m/s. On that particular sunspot, \cite[][]{Lop17} did not
observe converging flows towards the sunspot (L{\"o}ptien, private communication).

\section{Discussion and conclusions}

 We carried out CST and helioseismology investigation of solar differential rotation and meridian 
circulation at the Sun surface on a 30-day HMI/SDO sequence. We completed this study by applying the LCT and the FLCT onto the divergence field obtained from the CST. The usage of various methods allowed us to compare the results. 
 As a complementary study, we also examined the influence of a large sunspot on these large-scale flows with a specific 7-day HMI/SDO 
sequence. 

  We find that the large-scale flows measured by the CST on the solar surface and the same flow determined with the 
same data with the helioseismology in the first 1~Mm below the surface, are in good agreement in amplitude and 
direction. The torsional waves are also located at the same latitudes with amplitude of the same order. Using 
the CST method we are able to measure the meridional circulation correctly with only 3 days of data after averaging between $\pm 15\degr$ in longitude. This indicates to us the minimal number of days necessary to follow
correctly the evolution of the large-scale solar flows.

 The application of the LCT on the divergence maps does not show the same
differential rotation as the Doppler measurement. The FLCT applied also to the same data gives a result closer
  to the Doppler measurement but still shows lower amplitude for the high latitudes.
  Doppler is a direct determination of the velocities on the Sun, while LCT, FLCT, and CST are indirect
  velocity determination. In addition, LCT and FLCT do not measure the same flows as those determined by the Doppler.
  The LCT and FLCT are sensitive to the granules and intergranules horizontal displacements but Doppler
  is affected by its distance to the disc centre (vertical motion at the disc centre and horizontal motion
  elsewhere). Thus a detailed comparison of the large-scale flows obtained by the various methods
  (LCT, FLCT, CST, and Doppler) is still delicate and can reflect different properties of the turbulent flows of the photosphere.

 The differential rotation and meridional circulation  was determined by examining large flows in different activity
 regions: quiet Sun, sunspot, and plage. The differential rotation is sensitive to the relative separation motion between
 the leading sunspot and following plage. The meridional circulation is sensitive to the moat flow,  
 which clearly modifies its amplitude locally and probably to surrounding latitudes.

\begin{acknowledgements}
 The SDO data are courtesy of NASA and the SDO/HMI science team. 
This work was granted access to the HPC resources of CALMIP under the allocation 
2011-[P1115].  This work was supported by the CNRS Programme National 
Soleil Terre. M.~\v{S} is supported by the project RVO:67985815.The authors wish
to thank the anonymous referee for very helpful comments and suggestions that
improved the quality of the manuscript.
\end{acknowledgements}

\bibliographystyle{aa}
\bibliography{biblio}

\end{document}